\documentclass[aps,pra,reprint,superscriptaddress]{revtex4-2}

\usepackage{graphicx}
\usepackage{dcolumn}
\usepackage{bm}
\usepackage{amsmath,calc}
\usepackage{multirow,booktabs}
\usepackage{makecell}
\begin{document}


\title{Enhanced phase sensitivity in a Mach-Zehnder interferometer via photon recycling}


\author{Dong Li}
\affiliation{Microsystem and Terahertz Research Center, China Academy of Engineering Physics, Chengdu Sichuan 610200, P. R. China}
\affiliation{Institute of Electronic Engineering, China Academy of Engineering Physics, Mianyang Sichuan 621999, P. R. China}
\author{Chun-Hua Yuan}
\affiliation{State Key Laboratory of Precision Spectroscopy, Quantum Institute of Atom and Light, Department of Physics, East China Normal University, Shanghai 200062, China}
\author{Xiaoping Ma}
\affiliation{College of Mathematical and Physical Sciences, Qingdao University of Science and Technology, Qingdao, China}
\author{Qingle Wang}
\affiliation{School of Control and Computer Engineering, North China Electric Power University, Beijing 102206, China}
\affiliation{CAS Key Laboratory of Quantum Information, University of Science and Technology of China, Hefei 230026, China}
\author{Hwang Lee}
\affiliation{Hearne Institute for Theoretical Physics and Department of Physics and Astronomy, Louisiana State University, Baton Rouge, Louisiana 70803, USA}
\author{Yao Yao}
\email[]{yaoyao\_mtrc@caep.cn}
\affiliation{Microsystem and Terahertz Research Center, China Academy of Engineering Physics, Chengdu Sichuan 610200, P. R. China}
\affiliation{Institute of Electronic Engineering, China Academy of Engineering Physics, Mianyang Sichuan 621999, P. R. China}
\author{Weiping Zhang}
\affiliation{Department of Physics, Shanghai Jiao Tong University, and Tsung-Dao Lee Institute, Shanghai 200240, China}
\affiliation{Collaborative Innovation Center of Extreme Optics, Shanxi University, Taiyuan, Shanxi 030006, China}


\date{\today}

\begin{abstract}
We propose an alternative scheme for phase estimation in a Mach-Zehnder interferometer (MZI) with photon recycling. It is demonstrated that with the same coherent-state input and homodyne detection, our proposal possesses a phase sensitivity beyond the traditional MZI. For instance, it can achieve an enhancement factor of $\sim$9.32 in the phase sensitivity compared with the conventional scheme even with a photon loss of $10\%$ on the photon-recycled arm. From another point of view, the quantum Cram\'er-Rao bound (QCRB) is also investigated. It is found that our scheme is able to achieve a lower QCRB than the traditional one. Intriguingly, the QCRB of our scheme is dependent of the phase shift $\phi$ while the traditional scheme has a constant QCRB regardless of the phase shift. Finally, we present the underlying mechanisms behind the enhanced phase sensitivity (due to the improved number of photons inside the interferometer). We believe that our results may pave the way to enhance the phase sensitivity via photon recycling.
\end{abstract}


\maketitle

\section{Introduction}

With the development of quantum metrology, its theoretical and experimental progress has attracted increasing attention by scientists recently. Optical interferometer, a common tool for high-precision measurement, possesses a wide array of potential applications, including gravitational wave detection \cite{barsotti2018,mehmet2018}, optical lithography \cite{dowling2000}, optical gyroscope \cite{lefere1997fundamentals,fink2019entanglement,wu2020}. As depicted in Fig. \ref{fig1}(a), a typical Mach-Zehnder interferometer (MZI) consisting of two beam splitters works as follows: (i) an incident coherent-state beam is split into two modes by the first beam splitter; (ii) one mode experiences a relative phase shift $\phi$ while the other one retains as a reference; (iii) these two modes are then recombined at the second beam splitter. Finally, one extracts the expected information about the parameter $\phi$ by monitoring the output modes. Mathematically, the measurement quantity is regarded as an observable $A$. Accordingly, the phase sensitivity can be inferred from the variance of the observable $A$ via the linear error propagation formula $\Delta \phi = \langle \Delta A \rangle / |\partial \langle A \rangle/ \partial \phi|$.

As a matter of fact, the phase sensitivity in the traditional MZI is restricted by the shot-noise limit (SNL) \cite{caves1981}, $1/\sqrt{n}$, due to the shot noise (the quantum nature of light) where $n$ is the total incident mean photon number. To improve the phase sensitivity (even beyond the SNL), there usually exist two pathways: (i) decreasing $\langle \Delta A \rangle$ (i.e. reducing the quantum noise), which could be realized by injecting a beam with a sub-shot noise, for instance, squeezed states \cite{walls1983,walls2007,barnett2002,aasi2013}; (ii) increasing $|\partial \langle A \rangle/ \partial \phi|$ (i.e., enhancing the effective detection signal), which could be achieved by various methods, such as utilizing nonlinear beam splitter \cite{yurke1986,zhang2018super,zhang2021phase,ma2020enhanced,you2019conclusive,ma2018sub,gong2017intramode,gong2017phase,jiao2020,hu2018phase,jing2021} or using nonlinear phase shift \cite{woolley2008nonlinear,joo2012quantum,berrada2013quantum,cheng2014quantum,jiao2020nonlinear}.

In experiment, one can choose various measurement strategies, such as intensity detection \cite{marino2012effect,fritschel1992,regehr1995,mckenzie2002}, parity detection \cite{gerry2000heisenberg,seshadreesan2011parity,li2016phase,li2018effects}, and homodyne detection \cite{chen2016effects,kong2013experimental,hudelist2014quantum,li2014phase}. Particularly, we mainly focus on the homodyne detection (a measurement technique being used to monitor the quadrature of light). It is worth noting that homodyne detection is usually performed only on one output mode, while the other one is ignored in the traditional scheme \cite{hudelist2014quantum,li2014phase,pra2019} as shown in Fig. \ref{fig1}(a). Nevertheless, the discarded mode $b$ also contains the information about the parameter $\phi$. Naturally, we wonder whether it is able to enhance the phase sensitivity by reusing the ignored beam. Following this route, we propose an alternative scheme where the output mode $b$ is re-injected into the input port $b$ (so-called photon recycling) as shown in Fig. \ref{fig1}(b).

In fact, the technique of photon recycling has been theoretically proposed and experimentally realized in the Michelson interferometer (MI) \cite{a0prd1988,a0cqg2004,a0prd2003,a0ao1992,a0lrr2011,a0oe2021}. It has been proved that the photon recycling is an efficient technique for further signal increase, capitalizing both on an increase in circulating mean photon number and an increase in phase shift \cite{a0ao1992}. Nevertheless, the photon recycling is rarely applied in a MZI. From this point of view, it is also desired to explore the performance of the photon-recycled Mach-Zehnder interferometer.

In this work, we investigate the performance of this photon-recycled scheme from three related but distinct aspects: (i) the phase sensitivity via homodyne detection; (ii) the quantum Cram\'er-Rao bound (QCRB); (iii) the total mean photon number inside the interferometer (the photons experiencing and sensing the phase shift). First of all, it is demonstrated that this modified interferometer can realize an enhanced phase sensitivity compared with the traditional MZI. Second, we illustrate that this modified scheme can achieve a QCRB beyond the traditional one. In addition, unlike the traditional scheme where the QCRB is independent of the phase shift $\phi$, the QCRB of our scheme depends on $\phi$. Finally, we discuss the underlying mechanisms behind the enhanced phase sensitivity by analyzing the total mean photon number inside the interferometer.

This manuscript is organized as follows. In Sec. II, we briefly introduce the theoretical model of propagation of photons through this modified optical circuit. In Sec. III, we analyze the phase sensitivity via homodyne detection and QCRB in our scheme, and compare the performance between our scheme and the traditional one. In Sec. IV, we present the underlying mechanisms behind the enhanced phase sensitivity. Final remarks are given in Sec. V.

\section{Theoretical model}

\begin{figure}[tb]
\begin{center}
\includegraphics[width=0.5\textwidth]{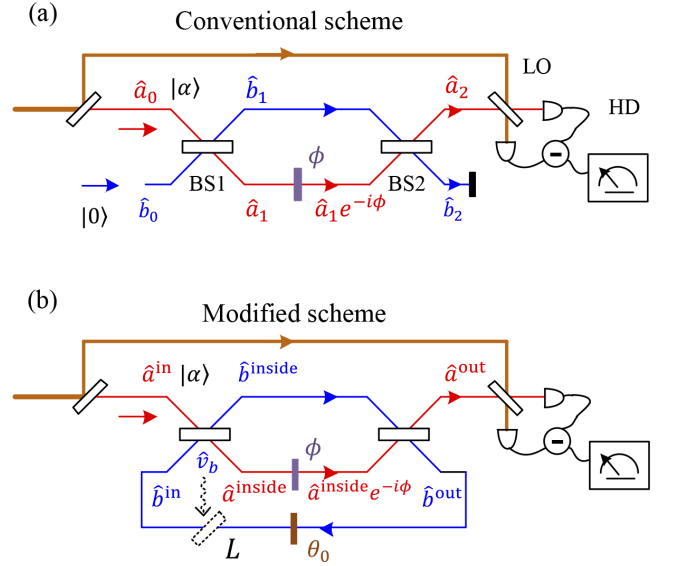}
\end{center}
\caption{Schemes for phase estimation: (a) the traditional MZI with the output beam $b$ being discarded, (b) the modified scheme with the output mode $b$ being re-injected into the input mode $b$. We have assumed that the output beam $b$ experiences a phase shift $\theta_0$ and photon loss $L$ before the re-injection. HD: homodyne detection, LO: local oscillator.}
\label{fig1}
\end{figure}

In contrast to the traditional scheme where the output beam $b$ is discarded (see Fig. \ref{fig1}(a)), we propose an alternative one where the output mode $b$ is reused via photon recycling (the output beam $b$ is re-injected into the input port $b$ as illustrated in Fig. \ref{fig1}(b)). In practical experiment, the photon loss is inevitable. Therefore, we consider the effect of photon loss induced by the light propagation in the recycling arm. In addition, we assume that the output beam $b$ experiences a phase shift $\theta_0$ before it is re-injected into the input port $b$.

In our scheme as depicted in Fig. \ref{fig1}(b), the input-output relation is found to be
\begin{align}
\begin{pmatrix}
\hat{a}^{\rm{out}} \\
\hat{b}^{\rm{out}} 
\end{pmatrix} &= S_{\rm{MZI}}
\begin{pmatrix}
\hat{a}^{\rm{in}} \\
\hat{b}^{\rm{in}} 
\end{pmatrix},
\label{eq001}
\end{align}

where $\hat{a}^{\rm{in}}$ ($\hat{a}^{\rm{out}}$) and $\hat{b}^{\rm{in}}$ ($\hat{b}^{\rm{out}}$) are the annihilation operators of input (output) modes. $S_{\rm{MZI}} = S_{\rm{BS}_2}S_{\phi}S_{\rm{BS}_1}$ represents the scattering matrix of a conventional MZI with 
\begin{align}
S_{\rm{BS}_1} = S_{\rm{BS}_2} = \frac{1}{\sqrt{2}}\begin{pmatrix}
1 & i \\
i & 1 \\
\end{pmatrix},
\end{align} and 
\begin{align}
S_{\phi} = \begin{pmatrix}
e^{-i\phi} & 0 \\
0 & 1 \\
\end{pmatrix}.
\end{align} 

Generally, the photon loss could be modeled by adding a fictitious beam splitter as shown in Fig. \ref{fig1}(b). Assume the photon-recycling arm with a loss rate $L$. After passing through the fictitious beam splitter, the mode transform of field $\hat{b}^{\rm{out}}$ is given by $\hat{b}^{\rm{out}\prime} = \sqrt{1-L} \hat{b}^{\rm{out}} + \sqrt{L} \hat{v}_{b}$ ($\hat{v}_{b}$ corresponds to the vacuum state). In the presence of phase shift $\theta_0$ and photon loss $L$, the re-injected mode $b$ could be expressed as 
\begin{align}
\hat{b}^{\rm{in}} = \sqrt{1-L} \hat{b}^{\rm{out}} e^{-i \theta_0} + \sqrt{L} \hat{v}_{b}.
\label{eqn1a}
\end{align}

Based on Eqs. (\ref{eq001}) and (\ref{eqn1a}), the output annihilation operators in Fig. \ref{fig1}(b) arrive at
\begin{align}
\hat{a}^{\rm{out}} = &(s_{11} + \frac{s_{12} s_{21} \sqrt{1-L} }{e^{i\theta_0} - s_{22}\sqrt{1-L}  }) \hat{a}^{\rm{in}} \nonumber\\
& + (1+\frac{s_{22}\sqrt{1-L} }{e^{i\theta_0} - s_{22} \sqrt{1-L} })s_{12}\sqrt{L}\hat{v}_b,
\label{aouteqn}\\
\hat{b}^{\rm{out}} = & \frac{s_{21}}{1 - s_{22} \sqrt{1-L} e^{-i\theta_0} } \hat{a}^{\rm{in}} +\frac{s_{22} \sqrt{L}}{1 - s_{22} \sqrt{1-L} e^{-i\theta_0} } \hat{v}_b,
\label{bouteqn}
\end{align}
where we have set
\begin{align}
S_{\rm{MZI}} = \begin{pmatrix}
s_{11} & s_{12} \\
s_{21} & s_{22} \\
\end{pmatrix},
\label{eq3}
\end{align}
with
\begin{align}
s_{11}&=\frac{1}{2}(e^{-i \phi} -1), \quad
s_{12}=\frac{i}{2}(e^{-i \phi} +1), \nonumber\\
s_{21}&=\frac{i}{2}(e^{-i \phi} +1), \quad
s_{22}=\frac{1}{2}(1-e^{-i \phi}).
\label{eq3}
\end{align}

Eq. (\ref{aouteqn}) is the output in our scheme in Fig. \ref{fig1}(b). To verify this result, we analyze our scheme via an alternative method and obtain the same expression. For the sake of clarity, the detailed analysis is shown in Appendix \ref{app000}. Since we monitor the final output mode $a$ in our scheme, Eq. (\ref{aouteqn}) would be used to estimate the phase shift $\phi$.

\section{phase sensitivity}

\subsection{Homodyne detection}

To extract the information of parameter $\phi$, we measure the quadrature of output mode $a$ via homodyne detection as show in Fig. \ref{fig1}(b). Mathematically, the output quadrature is defined as 
\begin{eqnarray}
\label{eqn_xa}
\hat{x}_{a} = \hat{a}^{\rm{out}\dagger} + \hat{a}^{\rm{out}},
\label{xaeqn}
\end{eqnarray}
where $\hat{a}^{\rm{out} \dagger}$ is the Hermitian conjugate of $\hat{a}^{\rm{out}}$. According to the linear error propagation formula, the phase sensitivity is given by
\begin{eqnarray}
\Delta \phi = \frac{\langle \Delta \hat{O} \rangle}{\left|\frac{\partial \langle\hat{O} \rangle}{\partial \phi}\right|},
\label{delta2}
\end{eqnarray}
where $\hat{O}$ is the operator of observable quantity and $\langle \Delta \hat{O} \rangle = \sqrt{\langle  (\hat{O})^2 \rangle -\langle  \hat{O} \rangle^2}$ is the corresponding variance. In homodyne detection, the measurement quantity is quadrature $\hat{O} = \hat{x}_a$. Accordingly, the phase sensitivity can be obtained by
\begin{align}
\Delta \phi = \frac{\langle \Delta \hat{x}_a \rangle}{\left|\frac{\partial \langle\hat{x}_a \rangle}{\partial \phi}\right|},
\label{delta3}
\end{align}
where it requires to calculate $\langle \hat{x}_a \rangle$ and $\langle \Delta \hat{x}_a \rangle$.

Consider a coherent state $|\alpha\rangle$ as input as shown in Fig. \ref{fig1}(b) (the complex number $\alpha = |\alpha|e^{i\theta_{\alpha}}$ denotes the amplitude). Without loss of generality, we set the irrelevant phase $\theta_{\alpha} = 0$. In this situation, by combining Eqs. (\ref{aouteqn}) and (\ref{xaeqn}), it is easily found that 
\begin{align}
\langle \hat{x}_a \rangle = \Upsilon \alpha + \Upsilon^{\ast} \alpha^{\ast}, \quad
\langle \Delta \hat{x}_a \rangle  = 1, 
\label{eqdelta}
\end{align}
where 
\begin{align}
\Upsilon=&\frac{e^{i\theta_0}(1- e^{i \phi})  - 2 \sqrt{1-L}}{2 e^{i (\phi+\theta_0)} + \sqrt{1-L} - e^{i \phi} \sqrt{1-L}},
\end{align}
$\alpha^{\ast}$ ($\Upsilon^{\ast}$) is the complex conjugate of $\alpha$ ($\Upsilon$), and the corresponding derivation is presented in Appendix \ref{app001}. If we suppose that $L=1$ (namely {\emph{block}} the photon-recycled arm), $\Upsilon=(e^{-i\phi}-1)/2$ is independent of $\theta_0$ (the phase shift on the photon-recycled arm) which sounds reasonable.

By inserting Eq. (\ref{eqdelta}) into Eq. (\ref{delta3}), the phase sensitivity yields

\begin{align}
\Delta \phi^{\rm{PR}} = \frac{1}{\Lambda_1} \Delta \phi^{\rm{Con}}_{\rm{SNL}},
\label{deltam}
\end{align}
where the superscript PR (Con) stands for the photon-recycling (conventional) scheme, $\Delta \phi^{\rm{Con}}_{\rm{SNL}} = 1/|\alpha|$ is the so-called SNL in the conventional scheme, and the explicit expression of $\Lambda_1$ is shown in Appendix \ref{appb}.

\begin{figure}[tb]
\begin{center}
\includegraphics[width=0.5\textwidth]{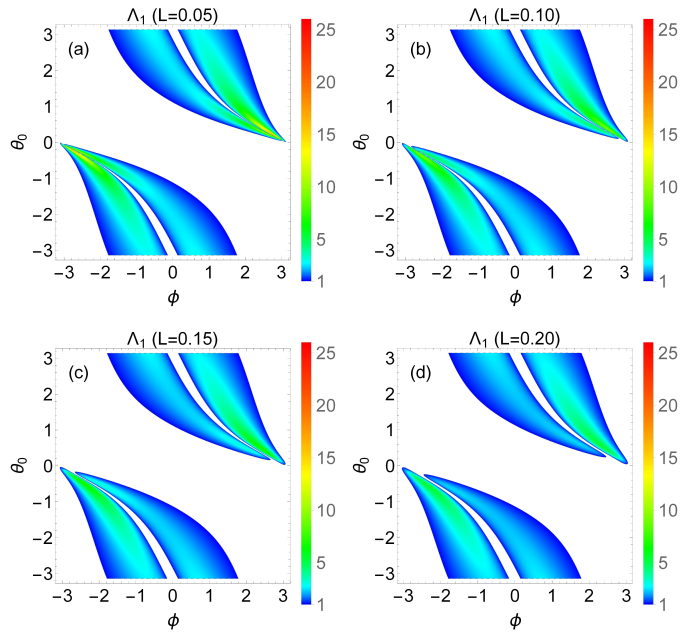}
\end{center}
\caption{The enhancement factor $\Lambda_{1}$ as a function of $\phi$ and $\theta_0$ in the presence of photon loss (a) $L=0.05$, (b) $L=0.10$, (c) $L=0.15$, and (d) $L=0.20$.}
\label{fig2}
\end{figure}

In fact, $\Lambda_1$ can be regarded as the enhancement factor of phase sensitivity. Fig. \ref{fig2} depicts the factor $\Lambda_1$ as a function of $\phi$ and $\theta_0$ with various losses $L=0.05, 0.10, 0.15,$ and $0.20$. It is shown that in the presence of moderate loss, the enhancement factor can reach above unity, $\Lambda_1>1$, in a {\emph{horn-shaped}} region which indicates that our scheme could achieve a phase sensitivity beyond the traditional one. It is worth pointing out that the maximum $\Lambda_1$ is achieved around the tip of the ``horn".

\begin{figure}[bt]
\begin{center}
\includegraphics[width=0.4\textwidth]{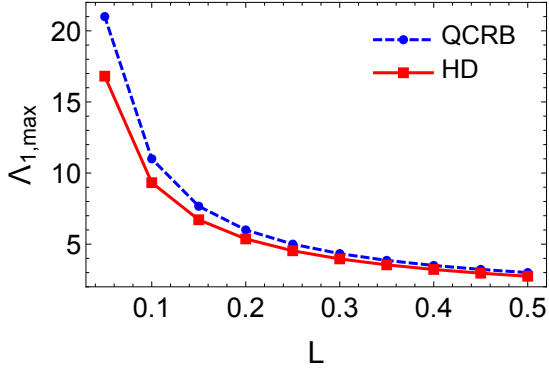}
\end{center}
\caption{The maximum enhancement factor $\Lambda_{1,\rm{max}}$ as a function of photon loss $L$. QCRB: quantum Cram\'er-Rao bound, HD: homodyne detection.}
\label{fig3}
\end{figure}

In an attempt to gain insight into the effect of loss on the minimum phase sensitivity, we quantitatively analyze the maximum $\Lambda_{1,\rm{max}}(= \Delta \phi_{\rm{SNL}}^{\rm{Con}}/\Delta \phi_{\rm{min}}^{\rm{PR}})$ from Eq. (\ref{eqn15}). Fig. \ref{fig3} plots $\Lambda_{1,\rm{max}}$ as a function of loss. It indicates that the minimum phase sensitivity increases with the increase of loss. Although the factor $\Lambda_{1,\rm{max}}$ decreases, it is still larger than unity with a moderate photon loss. In particular, when $L=0.10$, the maximum enhancement of phase sensitivity is roughly equal to $\Lambda_{1,\rm{max}} \simeq 9.32$ with $\theta_0\simeq0.3524$ rad and $\phi\simeq2.5702$ rad.

\subsection{Quantum Cram\'er-Rao bound}
\label{sec3}

Since the quantum Cram\'er-Rao bound (QCRB) \cite{helstrom1967minimum,helstrom1976quantum,braunstein1994statistical} gives the ultimate limit to precision, it is regarded as a reliable figure of merit when quantifying the performance of phase-estimation measurement scheme, which motivates us to investigate the QCRB in our scheme.

Theoretically, the output mode $a$ is always a pure Gaussian state in our analysis model. This is due to the facts: (i) in our proposal, the incident beam is a coherent state which is a pure Gaussian state; (ii) since the beam splitter acts as a Gaussian operation, the output is still a pure Gaussian state when a pure Gaussian state propagates through beam splitters (including the fictitious BS corresponding to the photon loss).

According to Ref. \cite{pinel2012}, the QCRB for a pure Gaussian state can be written as
\begin{align}
\Delta \phi_{\rm{QCRB}} = \left(\overline{X}'^{\top} \Gamma^{-1} \overline{X}' + \frac{\rm{tr(\Gamma' \Gamma^{-1})^2}}{4} \right)^{-1/2},
\label{eq10}
\end{align}
where the column vector of the expectation values of the quadratures $\overline{X}(= (\langle \hat{x}_{a}\rangle, \langle \hat{p}_{a}\rangle)^{\top})$ with $\hat{p}_{a} = i(\hat{a}^{\rm{out}\dagger} - \hat{a}^{\rm{out}})$, and the symmetrized covariance matrix
\begin{align}
\Gamma = \begin{pmatrix}\langle (\Delta \hat{x}_{a})^2\rangle&\langle\Delta(\hat{x}_a,\hat{p}_a)\rangle\\\langle\Delta(\hat{p}_a,\hat{x}_a)\rangle&\langle (\Delta \hat{p}_{a})^2\rangle \end{pmatrix},
\end{align}
with $\langle\Delta(\hat{O}_1,\hat{O}_2)\rangle = \frac{1}{2}\langle  \hat{O}_{1} \hat{O}_{2} +  \hat{O}_{2} \hat{O}_{1}\rangle - \langle\hat{O}_{1} \rangle \langle\hat{O}_{2} \rangle$ ($\hat{O}_1,\hat{O}_2 = \hat{x}_a,\hat{p}_a$), $O'=\partial O/\partial \phi$, and $O^{\top}$ ($O^{-1}$) is the transpose (inverse) of $O$.

In this modified scheme, $\overline{X}$ and $\Gamma$ are found to be
\begin{align}
\overline{X} = \big(2{\rm{Re}}(\Upsilon \alpha), 2{\rm{Im}}(\Upsilon \alpha)\big)^{\top}, \quad
\Gamma = 
\begin{pmatrix}
1&0\\
0&1
\end{pmatrix},
\label{eq11}
\end{align}
where Re($\Upsilon\alpha$) (Im($\Upsilon\alpha$)) is the real (imaginary) part of $\Upsilon\alpha$.

\begin{figure}[tb]
\begin{center}
\includegraphics[width=0.5\textwidth]{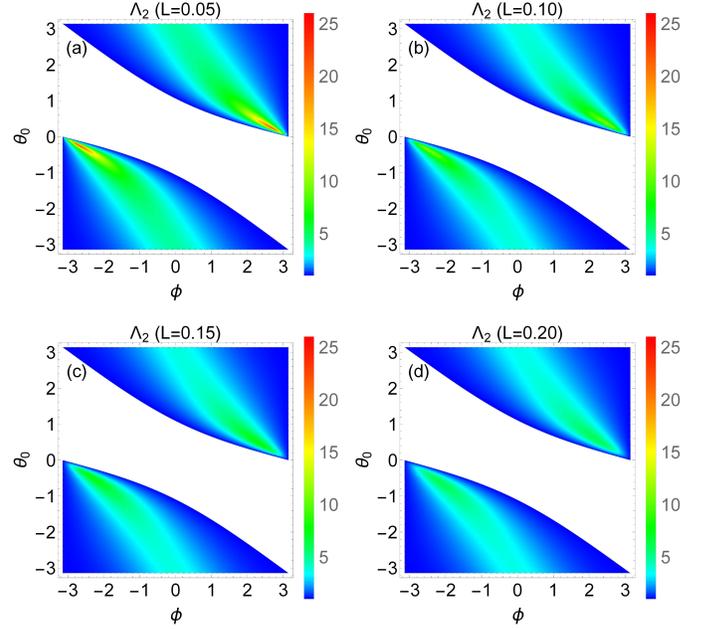}
\end{center}
\caption{The enhancement factor $\Lambda_2$ as a function of $\phi$ and $\theta_0$ with photon loss (a) $L=0.05$, (b) $L=0.10$, (c) $L=0.15$, and (d) $L=0.20$.}
\label{fig4}
\end{figure}

By substituting Eq. (\ref{eq11}) into Eq. (\ref{eq10}), the QCRB is cast into
\begin{align}
\Delta \phi_{\rm{QCRB}}^{\rm{PR}} = \frac{1}{\Lambda_2} \Delta \phi_{\rm{QCRB}}^{\rm{Con}},
\end{align}
where $\Delta \phi_{\rm{QCRB}}^{\rm{Con}} = 1/|\alpha|$ is the QCRB of the traditional scheme and the explicit expression of $\Lambda_2$ is present in Appendix \ref{appb}. Fig. \ref{fig4} shows $\Lambda_2$ as a function of $\phi$ and $\theta_0$ with different loss $L=0.05,0.10,0.15$ and $0.20$. Similar to $\Lambda_1$, $\Lambda_2$ also reaches above unity in a {\emph{horn-shaped}} region and the maximum $\Lambda_2$ is achieved around the tip of the ``horn" as well.

According to Eq. (\ref{eqn20}), one can numerically compute the maximum $\Lambda_{\rm{2,\rm{max}}}$. As shown in Fig. \ref{fig3}, with the increase of loss, the factor $\Lambda_{\rm{2,\rm{max}}}$ decreases, which yields that as the loss increases, the minimum QCRB becomes worse. Although $\Lambda_{\rm{2,\rm{max}}}$ decreases, it is still larger than one. That is to say, with a moderate loss, our scheme can still achieve a QCRB beyond the tradition scheme. It is worth noting that in contrast to the traditional MZI where the QCRB is independent of $\phi$, the QCRB in our scheme depends on $\phi$.

\section{discussion}
We would like to discuss the underlying mechanisms behind the enhanced phase sensitivity. Let us consider the total mean photon number, $\hat{n}_{\rm{T}}$, inside the interferometer where $\hat{n}_{\rm{T}}$ is defined as
\begin{align}
\langle \hat{n}_{\rm{T}} \rangle = \langle \hat{n}_{a}^{\rm{inside}} \rangle + \langle \hat{n}_{b}^{\rm{inside}} \rangle,
\label{eqn23}
\end{align}
with $\hat{n}_{j}^{\rm{inside}} = \hat{j}^{\rm{inside} \dagger} \hat{j}^{\rm{inside}}$ ($j = a,b$) ($\hat{j}^{\rm{inside}}$ corresponds to the mode $j$ inside the interferometer as shown in Fig. \ref{fig1}(b)). The total mean photon number is found to be
\begin{align}
\langle \hat{n}_{\rm{T}} \rangle & =\Lambda_3|\alpha|^2,
\label{eq17}
\end{align}
where the explicit expression of $\Lambda_3$ is shown in Appendix \ref{app002}.

\begin{figure}[tb]
\begin{center}
\includegraphics[width=0.5\textwidth]{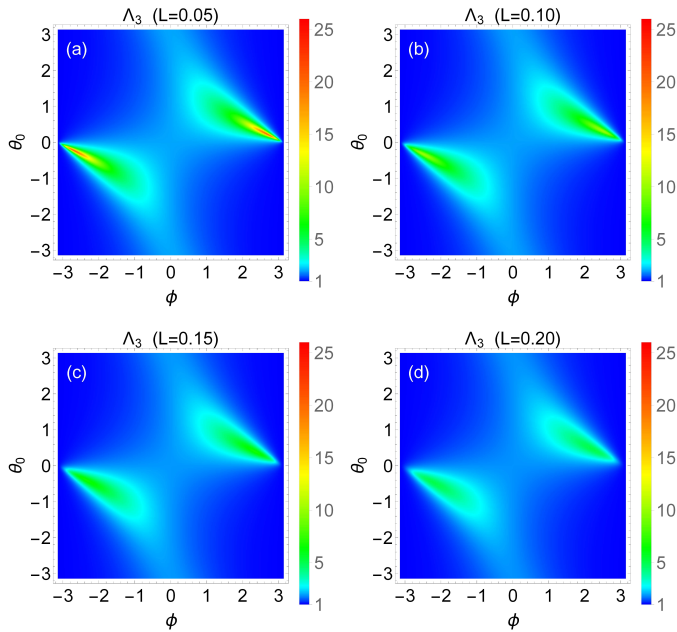}
\end{center}
\caption{The factor $\Lambda_3$ as a function of $\phi$ and $\theta_0$ with loss (a) $L=0.05$, (b) $L=0.10$, (c) $L=0.15$, and (d) $L=0.20$.}
\label{fig5}
\end{figure}

Fig. \ref{fig5} illustrates $\Lambda_3$ as a function of $\phi$ and $\theta_0$ with various losses $L=0.05$, $0.10$, $0.15$ and $0.20$. It is easy to check that $\Lambda_3 \ge 1$ is always valid, which indicates that our scheme possesses more photons inside the interferometer than the conventional one. In fact, photons inside the interferometer are usually regarded as the essential resource to experience and sense the relative phase shift. Intuitively, the more resource we use, the better phase sensitivity it can achieve \cite{taylor2016quantum}. Therefore, the phase sensitivity is improved in this modified scheme. It is worth noting that the mean photon number inside the interferometer $\langle \hat{n}_{\rm{T}}\rangle$ depends on $\phi$, leading to the phenomenon of QCRB being dependent of $\phi$ in our scheme. The photon-recycled Michelson interferometers have been proposed. Appendix \ref{appcomp} compares our scheme with these previous light-recycled ones. It is found that our scheme possesses a different structure of optical circuit.

\section{conclusion}
In summary we propose an alternative scheme for phase estimation in a MZI with photon recycling. We investigate the performances of our scheme, including the phase sensitivity via homodyne detection and quantum Cram\'er-Rao bound. It is demonstrated that this modified scheme is able to achieve an enhanced performance in contrast to the traditional MZI. Moreover, we present the physical mechanisms behind the improved phase sensitivity. We believe that our scheme may offer another route to implement a high-precision phase estimation exploiting photon recycling.

\section*{acknowledgments}
This research is supported by the National Natural Science Foundation of China (NSFC) (Grants No. 12104423, No. 12175204, No. 61875178, No. 62004180).

\appendix
\section{An alternative method to analyze the optical circuit in our scheme}
\label{app000}

\begin{figure}[tbh]
\begin{center}
\includegraphics[width=0.5\textwidth]{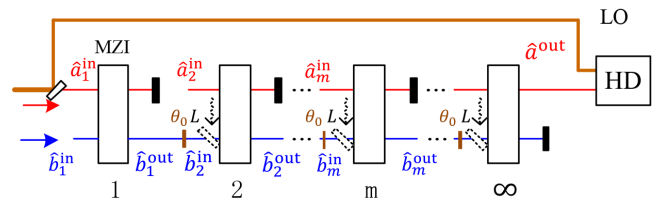}
\end{center}
\caption{The iterative-structure of series of MZIs where $m$-th output mode $b$ is injected into the $(m+1)$-th input port $b$. This model is equivalent to the scheme in Fig. \ref{fig1}(b).}
\label{fig10}
\end{figure}

To verify the result of Eq. (\ref{aouteqn}), we provide an alternative method to analyze our scheme. In fact, the optical circuit in Fig. \ref{fig1}(b) is equivalent to a series of conventional MZIs with an iterative structure (see Fig. \ref{fig10}), where each output beam $b$ is injected into the following input port $b$. This iterative-structure optical system has an input-output relation
\begin{align}
\begin{pmatrix}
\hat{a}_1^{\rm{out}} \\
\hat{b}_1^{\rm{out}} 
\end{pmatrix} &= S_{\rm{MZI}}
\begin{pmatrix}
\hat{a}_1^{\rm{in}} \\
\hat{b}_1^{\rm{in}} 
\end{pmatrix},\nonumber\\
\begin{pmatrix}
\hat{a}_2^{\rm{out}} \\
\hat{b}_2^{\rm{out}} 
\end{pmatrix} &= S_{\rm{MZI}}
\begin{pmatrix}
\hat{a}_2^{\rm{in}} \\
\hat{b}_2^{\rm{in}} 
\end{pmatrix},\nonumber\\
&\mathrel{\makebox[\widthof{=}]{\vdots}} \nonumber\\
\begin{pmatrix}
\hat{a}_m^{\rm{out}} \\
\hat{b}_m^{\rm{out}} 
\end{pmatrix} &= S_{\rm{MZI}}
\begin{pmatrix}
\hat{a}_m^{\rm{in}} \\
\hat{b}_m^{\rm{in}} 
\end{pmatrix},\nonumber\\
&\mathrel{\makebox[\widthof{=}]{\vdots}} 
\label{eq4a}
\end{align}
where $\hat{O}^{\rm{in}}_k$ and $\hat{O}^{\rm{out}}_k$ ($O=a,b$; $k=1,2,\cdots$) are the annihilation operators of input and output modes in the $k$-th MZI, respectively. 

Before re-injection, the mode $b$ experiences the phase shift $\theta_0$ and photon loss $L$. Therefore, the input mode $b$ in $(k+1)$-th MZI is found to be 
\begin{align}
\hat{b}_{k+1}^{\rm{in}} = \sqrt{1-L} \hat{b}_{k}^{\rm{out}} e^{-i \theta_0} + \sqrt{L} \hat{v}_{b},
\label{eqn1a}
\end{align}
where $\hat{b}_{k}^{\rm{out}}$ ($k=1,2,\cdots$) represents the output mode $b$ in the $k$-th MZI, and $\hat{v}_{b}$ denotes the vacuum.

Based on Eqs. (\ref{eq4a}) and (\ref{eqn1a}), the output annihilation operators of the $m$-th MZI in Fig. \ref{fig10} arrive at
\begin{align}
\hat{a}_m^{\rm{out}} =&s_{11} \hat{a}_{1}^{\rm{in}} + s_{12} \gamma^{m-1} \hat{b}_1^{\rm{in}} +s_{12} \frac{1-\gamma^{m-1}}{1-\gamma} \kappa, \label{eq2a}\\
\hat{b}_m^{\rm{out}} =&s_{21}  \hat{a}_{1}^{\rm{in}} + s_{22} \gamma^{m-1} \hat{b}_1^{\rm{in}} + s_{22} \frac{1-\gamma^{m-1}}{1-\gamma} \kappa,
\label{eq2b}
\end{align}
where we have set 
\begin{align}
\gamma = &\sqrt{1-L} e^{-i\theta_0}s_{22},\nonumber\\
\kappa = &\sqrt{1-L} e^{-i\theta_0}s_{21} \hat{a}_1^{\rm{in}} + \sqrt{L} \hat{v}_b.
\end{align}

Letting $m$ tend to infinity ($m \to \infty$) in Eqs. (\ref{eq2a}) and (\ref{eq2b}), one can obtain the output annihilation operators of modes $a$ and $b$ as shown in Fig. \ref{fig10}, 
\begin{align}
\hat{a}^{\rm{out}}  =&  \hat{a}^{\rm{out}}_{m}|_{m \to \infty} \nonumber\\
=& (s_{11} + \frac{s_{12} s_{21} \sqrt{1-L} }{e^{i\theta_0} - s_{22}\sqrt{1-L}  }) \hat{a}_1^{\rm{in}} \nonumber\\
& + (1+\frac{s_{22}\sqrt{1-L} }{e^{i\theta_0} - s_{22} \sqrt{1-L} })s_{12}\sqrt{L}\hat{v}_b,
\label{eq5aa}\\
\hat{b}^{\rm{out}} =& \hat{b}^{\rm{out}}_{m}|_{m \to \infty}\nonumber\\
=&  \frac{s_{21}}{1 - s_{22} \sqrt{1-L} e^{-i\theta_0} } \hat{a}_1^{\rm{in}} +\frac{s_{22} \sqrt{L}}{1 - s_{22} \sqrt{1-L} e^{-i\theta_0} } \hat{v}_b.
\label{eq5bb}
\end{align}
It is easily found that Eq. (\ref{eq5aa}) is the same as Eq. (\ref{aouteqn}) where we arrive at the same result via two different methods.

\section{Derivation of the expectation values $\langle \hat{x}_a \rangle$ and $\langle \Delta \hat{x}_a \rangle$}
\label{app001}

The expectation value of output quadrature is given by
\begin{align}
\langle \hat{x}_a \rangle = \langle \hat{a}^{\rm{out}\dagger}+ \hat{a}^{\rm{out}} \rangle,
\label{xa10}
\end{align}

By inserting Eq. (\ref{aouteqn}) into Eq. (\ref{xa10}), one can rewrite Eq. (\ref{xa10}) as
\begin{align}
\langle \hat{x}_a\rangle = \Upsilon \alpha +\Upsilon^{\ast} \alpha^{\ast},
\label{eqnxa00}
\end{align}
where $O^{\ast}$ ($O=\Upsilon,\alpha$) is the complex conjugate of $O$ and
\begin{align}
\Upsilon=&\frac{e^{i\theta_0} - e^{i (\phi+\theta_0)} - 2 \sqrt{1-L}}{2 e^{i (\phi+\theta_0)} + \sqrt{1-L} - e^{i \phi} \sqrt{1-L}},
\label{infty}
\end{align}

The fluctuation $\langle \Delta \hat{x}_a \rangle$ is defined as
\begin{align}
\langle \Delta \hat{x}_a \rangle \equiv \sqrt{\langle (\hat{x}_a)^2\rangle - \langle \hat{x}_a \rangle^2},
\label{xa00}
\end{align}
where
\begin{align}
\langle (\hat{x}_a)^2\rangle = \langle (\hat{a}^{\rm{out}\dagger}+\hat{a}^{\rm{out}})^2\rangle.
\label{xa2eqna1}
\end{align}
According to Eq. (\ref{aouteqn}), Eq. (\ref{xa2eqna1}) could be recast as
\begin{align}
\langle (\hat{x}_a)^2\rangle = (\Upsilon \alpha +\Upsilon^{\ast} \alpha^{\ast})^2 + 1.
\label{eqnxa200}
\end{align}

Substituting Eqs. (\ref{eqnxa00}), (\ref{infty}), and (\ref{eqnxa200}) into Eq. (\ref{xa00}), one can obtain
\begin{align}
\langle \Delta \hat{x}_a \rangle =1.
\label{xa01}
\end{align}

\section{$\Lambda_1$ and $\Lambda_2$}
\label{appb}

The explicit expression of $\Lambda_1$ is found to be
\begin{widetext}
\begin{align}
\Lambda_1 = 4 \left|\frac{\left(\sqrt{1-L} \cos\theta_0 -1\right)\left[\left(2-L-2 \sqrt{1-L}\cos \theta_0 \right) \sin\phi-\sqrt{1-L} (\cos \phi+1)\sin \theta_0\right]}{[-L-(1-L) \cos\phi-2 \sqrt{1-L} \cos \theta_0+2 \sqrt{1-L} \cos (\phi+\theta_0)+3]^2}\right|,
\label{eqn15}
\end{align}
\end{widetext}

and $\Lambda_2$ is given by
\begin{align}
\Lambda_2 =  \left|\frac{2\left(-2 \sqrt{1-L} \cos
   \theta_0-L+2\right)}{2
   \sqrt{1-L} \Theta-(1-L) \cos
   \phi -L+3}\right|,
   \label{eqn20}
\end{align}
where 
\begin{align}
\Theta= \cos \left(\theta _0+\phi
   \right)- \cos
   \theta _0.
   \label{eqn20a}
\end{align}

\section{Derivation of the total mean number of photon inside the interferometer $\langle \hat{n}_{\rm{T}} \rangle$}
\label{app002}

\begin{figure}[h]
\begin{center}
\includegraphics[width=0.5\textwidth]{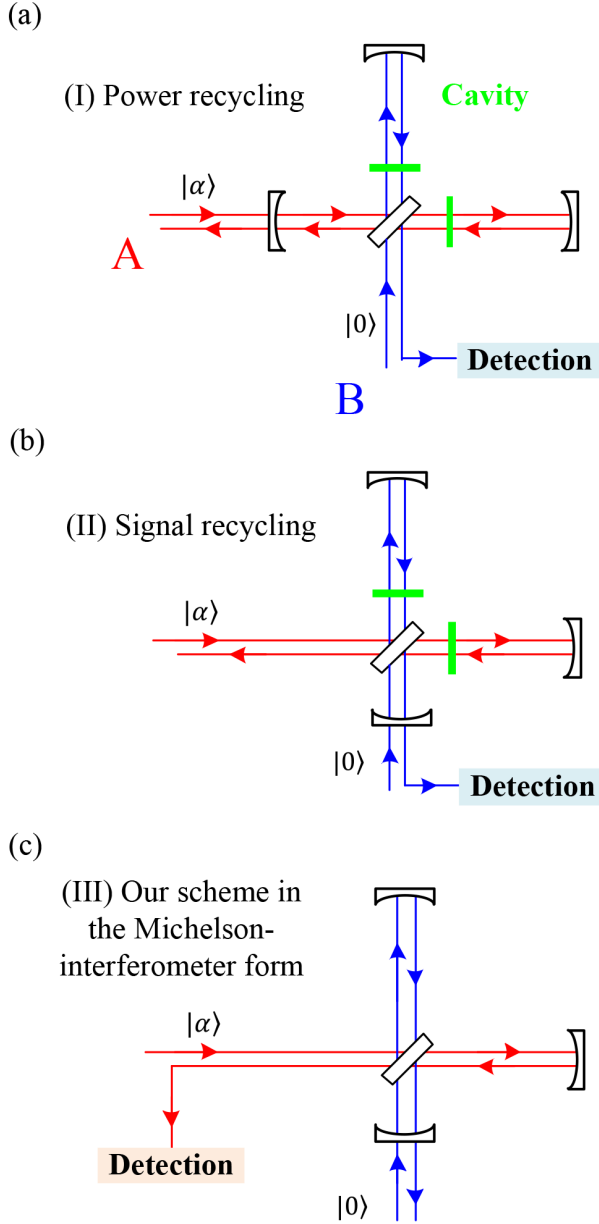}
\end{center}
\caption{Comparison between different photon-recycled schemes: (a) the power-recycled Michelson interferometer, (b) the signal-recycled Michelson interferometer, (c) our scheme in a Michelson-interferometer form. In (a) and (b), the green squares denote the mirrors.}
\label{fig6}
\end{figure}

Due to the energy conservation for photons propagating through a linear beam splitter, it is easy to verify that the total mean number of output photon is equal to the mean number of photon inside the interferometer 
\begin{align}
\langle \hat{n}_{a}^{\rm{inside}} \rangle + \langle \hat{n}_{b}^{\rm{inside}} \rangle=\langle \hat{n}_{a}^{\rm{out}} \rangle + \langle \hat{n}_{b}^{\rm{out}} \rangle,
\label{eqn24}
\end{align}
where $\hat{n}_{j}^{\rm{out}} \equiv \hat{j}^{\rm{out} \dagger} \hat{j}^{\rm{out}}$ ($j = a,b$). By inserting Eq. (\ref{eqn24}) into Eq. (\ref{eqn23}), the total mean photon number can be rewritten as
\begin{align}
\langle \hat{n}_{\rm{T}} \rangle &= \langle \hat{n}_{a}^{\rm{out}} \rangle + \langle \hat{n}_{b}^{\rm{out}} \rangle\nonumber\\
&=\langle \hat{a}^{\rm{out}\dagger} \hat{a}^{\rm{out}} \rangle + \langle \hat{b}^{\rm{out}\dagger} \hat{b}^{\rm{out}} \rangle.
\label{eq17abc}
\end{align}

From Eq. (\ref{aouteqn}), the mean photon number of output mode $a$ is given by
\begin{align}
\langle \hat{a}^{\rm{out}\dagger} \hat{a}^{\rm{out}} \rangle = |\Upsilon|^2 |\alpha|^2,
\label{naa01}
\end{align}
where $\Upsilon$ is present in Eq. (\ref{infty}).

Similarly, according to Eq. (\ref{bouteqn}), one can obtain
\begin{align}
\langle \hat{b}^{\rm{out}\dagger} \hat{b}^{\rm{out}} \rangle = |\Xi|^2 |\alpha|^2,
\label{nbb01}
\end{align}
where
\begin{align}
\Xi=\frac{i e^{i \theta _0} (1+e^{i \phi })}{2 e^{i
   (\theta _0+\phi )}-\sqrt{1-L} e^{i \phi
   }+\sqrt{1-L}}.
\label{nyy01}
\end{align}

The total mean photon number is then found to be
\begin{align}
\langle \hat{n}_{\rm{T}} \rangle = \Lambda_3 |\alpha|^2,
\label{ntot01}
\end{align}
where
\begin{align}
\Lambda_3=&|\Upsilon|^2+|\Xi|^2 \nonumber\\
=&\frac{2 \sqrt{1-L} \Theta-(1-L)
   \cos \phi -2L
   +4}{2 \sqrt{1-L} \Theta-(1-L)
   \cos \phi -L+3},
   \label{eqnm3}
\end{align}
with $\Theta$ being defined in Eq. (\ref{eqn20a}).

\section{Comparison with other light-recycled schemes}
\label{appcomp}

The photon recycling is a mature technique which has been proposed and experimentally realized in the Michelson interferometer, for instance, the power-recycled Michelson interferometer (Fig. \ref{fig6}(a)) and the signal-recycled Michelson interferometer (Fig. \ref{fig6}(b)) \cite{a0cqg2004}. Since the MZI is equivalent to the MI where two beam splitters in the MZI can be reduced to one in the MI. From this point of view, our scheme (light-recycled MZI) is equivalent to the light-recycled MI as shown in Fig. \ref{fig6}(c). For further understanding, the comparison between our scheme and the conventional light-recycled ones is performed. Although the light recycling is commonly utilized in these three schemes, there still exists obvious difference: 

(i)  Different optical circuit in each arm. Note that a Fabry-Perot cavity is placed in each interferometer arm in the conventional schemes but does not in our case.

(ii) Different detection port. The incident beams are injected from the same port (port A) in these three schemes, but the detection ports are different where ours is at port A while the others at port B.



%

\end{document}